\documentclass[12pt]{article}

\ifx\pdfoutput\undefined
\usepackage[dvips,bookmarks]{hyperref}
\else
\usepackage{hyperref}
\fi
\hypersetup{colorlinks=false,bookmarksopen,bookmarksnumbered,citecolor=blue,
   pdfstartview=FitH}

\usepackage{latexsym}
\usepackage{amssymb,amsfonts,amsmath}
\usepackage{graphicx} 
\usepackage{indentfirst}
\usepackage{bbm}
\usepackage{amssymb}
\usepackage{verbatim}
\usepackage{amsmath, amsthm,amssymb}
\usepackage{mathrsfs}
\usepackage{hyperref}
\usepackage{amsfonts}
\usepackage{dsfont}
\usepackage{slashed, tensor}
\usepackage{booktabs}
\usepackage{graphicx}
\usepackage{mathrsfs}
\parskip=\medskipamount

\newcommand {\cA}{{\cal A}}
\newcommand {\cB}{{\cal B}}
\newcommand {\cC}{{\cal C}}
\newcommand {\cD}{{\cal D}}
\newcommand {\cE}{{\cal E}}
\newcommand {\cF}{{\cal F}}
\newcommand {\cG}{{\cal G}}

\newcommand {\cL}{{\cal L}}
\newcommand {\cM}{{\cal M}}
\newcommand {\cN}{{\cal N}}

\newcommand {\cR}{{\cal R}}

\newcommand {\cW}{{\cal W}}

\def\a{\alpha}
\def\b{\beta}
\def\c{\chi}
\def\d{\delta}

\def\f{\phi}
\def\g{\gamma}
\def\G{\Gamma}

\def\q{\theta}
\def\r{\rho}
\def\s{\sigma}

\def\D{\Delta}

\def\J{\Psi}
\def\L{\Lambda}

\def\S{\Sigma}

\def\X{\Xi}
\def\ri{{\rm i}}

\newcommand{\ad}{{\dot{\alpha}}}                           %new
\newcommand{\bd}{{\dot{\beta}}}                            %new
\newcommand{\ve}{\varepsilon}                            %new
\newcommand{\cDB}{{\bar\cD}}                            %new

\newcommand{\hf}{\frac12}
\newcommand{\be}{\begin{equation}}
\newcommand{\ee}{\end{equation}}
\newcommand{\bea}{\begin{eqnarray}}
\newcommand{\eea}{\end{eqnarray}}
\newcommand{\non}{\nonumber}
\newcommand{\ba}{\begin{array}}
\newcommand{\ea}{\end{array}}

\newcommand{\1}{{\underline{1}}}
\newcommand{\2}{{\underline{2}}}

\newcommand{\dsC}{{\mathbb C}}

\newcommand{\bm}[1]{\mbox{\boldmath$#1$}}

\def\double #1{#1{\hbox{\kern-2pt $#1$}}}

\newcommand{\gd}{{\dot\g}}
\newcommand{\dd}{{\dot\d}}

\newcommand{\sba}{{\bar{\s}}}

\newcommand{\bsubeq}{\begin{subequations}}
\newcommand{\esubeq}{\end{subequations}}

\newcommand{\eps}{{\ve}}

\newcommand{\dalpha}{{\dot{\alpha}}}
\newcommand{\dbeta}{{\dot{\beta}}}

\newcommand{\eol}{\notag \\}
\newcommand{\rd}{\mathrm d}
\numberwithin{equation}{section}
\begin{document}
%%%%%%%%%%%%%%%%
%%%%%%%%%%%%%%%%

\begin{titlepage}
\begin{flushright}
July, 2015 \\
Revised version: August, 2015\\
\end{flushright}
\vspace{5mm}

\begin{center}
{\Large \bf 
On curvature squared terms in \mbox{$\bm{ \cN=2}$} supergravity
}
\\ 
\end{center}

\begin{center}

{\bf
Sergei M. Kuzenko${}^{a}$, Joseph Novak${}^{b}$
} \\
\vspace{5mm}

\footnotesize{
${}^{a}${\it School of Physics M013, The University of Western Australia\\
35 Stirling Highway, Crawley W.A. 6009, Australia}}  
~\\
\vspace{2mm}
\footnotesize{
${}^{b}${\it Max-Planck-Institut f\"ur Gravitationsphysik, Albert-Einstein-Institut,\\
Am M\"uhlenberg 1, D-14476 Golm, Germany.}
}
\vspace{2mm}
~\\
\texttt{sergei.kuzenko@uwa.edu.au, joseph.novak@aei.mpg.de}\\
\vspace{2mm}

\end{center}

\begin{abstract}
\baselineskip=14pt

We present the $\cN = 2$ supersymmetric completion
of a scalar curvature squared term in a completely 
gauge independent form. We also elaborate on its 
component structure.

\end{abstract}

\vfill

\vfill
\end{titlepage}

\newpage
\renewcommand{\thefootnote}{\arabic{footnote}}
\setcounter{footnote}{0}

%%%%%%%%%%%%%%%%%%%%%%%%%%%%%%%%%%%%%%%%%%%%%%%%%%%%%%
%%%%%%%%%%%%%%%%%%%%%%%%%%%%%%%%%%%%%%%%%%%%%%%%%%%%%%

\allowdisplaybreaks

\section{Introduction}

Recently, there has been renewed interest in $R^2$ gravity 
\cite{Kounnas:2014gda,Kehagias:2015ata,Alvarez-Gaume:2015rwa} 
and $\cN=1$ supergravity \cite{Ferrara:2015ela}. 
In particular, it has been confirmed that the pure $R^2$ gravity theory is 
ghost free \cite{Alvarez-Gaume:2015rwa}. This provides a rationale
to look more closely at the structure of curvature squared terms in 
four-dimensional (4D) $\cN=2$ supergravity. 
On dimensional grounds, all such terms should be given by chiral integrals.

The $\cN=2$ locally supersymmetric invariant 
$I_{C^2_{abcd}}$ 
containing the Weyl tensor squared
(which coincides with the action for $\cN=2$ conformal supergravity) 
was constructed by Bergshoeff, de Roo and de Wit 
almost thirty five years ago \cite{BdRdW}. 
However, the $\cN=2$ supersymmetric extension $I_{R^2_{ab} -\frac{1}{3} R^2}$
of the term  $R^{ab}R_{ab} -\frac{1}{3} R^2 $ was obtained only two years ago 
\cite{BdeWKL}. A special combination of the super-Weyl
invariants  $I_{C^2_{abcd}}$ and 
$I_{R^2_{ab} -\frac{1}{3} R^2}$ constitutes the $\cN=2$ Gauss-Bonnet term 
\cite{BdeWKL}. In this note we describe a third curvature squared 
invariant -- a locally supersymmetric extension of the $R^2$ term that is of special interest in the context 
of the ideas advocated in \cite{Kounnas:2014gda,Kehagias:2015ata,Alvarez-Gaume:2015rwa,Ferrara:2015ela}. 
Although the invariant has been discussed in \cite{Ketov, CeresoleSearch}, there has not appeared 
a complete description of the invariant
 due to some missing elements. In particular, the invariant was 
 given in \cite{Ketov} in a special gauge and the explicit component action has never 
 been worked out. 
In this note we provide a more complete exposition of the invariant and construct it 
in a gauge independent form.
 
This note is organised as follows. In section \ref{section2} we present the superspace 
description of curvature squared invariants within $\cN =2$ superspace. In section \ref{section3} 
we elaborate on the component structure of a $\cN = 2$ supersymmetric invariant containing 
a curvature squared term. Section \ref{section4} is devoted to a discussion of our results.

We have included a couple of technical appendices. Appendix \ref{AppendA} 
provides the essential details of the formulation for $\cN = 2$ conformal supergravity \cite{KLRT-M1}
in $\rm SU(2)$ superspace \cite{Grimm}, while Appendix \ref{confSuperspace} 
summarises the important details of conformal superspace \cite{Butter11}.

%%%%%%%%%%%%%%%%%%%%%%%%%%%%%%%%%%%%%%%%%%%%%%%%%%%%%%
%%%%%%%%%%%%%%%%%%%%%%%%%%%%%%%%%%%%%%%%%%%%%%%%%%%%%%

\section{The curvature squared invariants in superspace} \label{section2}

In this section we use the formulation for $\cN=2$ conformal 
supergravity  \cite{KLRT-M1} in SU(2) superspace \cite{Grimm}.
Some technical details concerning this supergravity formulation
are collected in Appendix \ref{AppendA}.
We proceed by recalling the explicit structure of the 
invariants  $I_{C^2_{abcd}}$ and 
$I_{R^2_{ab} -\frac{1}{3} R^2}$. 

The invariant containing the Weyl tensor squared is
\bea
I_{C^2_{abcd}} =  \int \rd^4x\, \rd^4\q\, \cE\,  W^{\a\b}W_{\a\b} 
~+~{\rm c.c.} \ ,
\label{2.1}
\eea
where $W_{\a\b}$ is the super-Weyl tensor, see Appendix \ref{AppendA}, 
and $\cE$ is the chiral density, see, e.g., \cite{KT-M09} for the definition of 
$\cE$.

The invariant containing $R^{ab}R_{ab} -\frac{1}{3} R^2 $ is 
\bea
I_{R^2_{ab} -\frac{1}{3} R^2}= \int \rd^4x\, \rd^4\q\, \cE\,  \X ~+~{\rm c.c.} ~,
\label{2.2}
\eea
where  $\X$  denotes the following composite scalar \cite{BdeWKL}:
\bea
\X :=  \frac{1}{6}  \bar{\cD}^{ij} \bar S_{ij}+  \bar S^{ij} \bar S_{ij}
+ \bar Y_{\dalpha \dbeta} \bar Y^{\dalpha \dbeta}~, \qquad
\cDB_{ij}:=\cDB_{\ad(i}\cDB_{j)}^\ad~.
\eea   
The torsion superfields $S_{ij}$, $W_{\a\b}$ and $Y_{\a\b}$ and their conjugates
$\bar S^{ij} $, $\bar W_{\ad \bd}$  and $\bar Y_{\ad \bd}$ 
are defined in Appendix \ref{AppendA}.
The fundamental properties of $\X$ are as follows \cite{BdeWKL}: \\
(i) it  is covariantly chiral, 
\bea
\bar \cD^\ad_i \X=0~;
\eea
 (ii) its super-Weyl transformation  is  
 \bea
 \d_\s \X = 2\s \X -2 \bar \D \bar  \s ~.
\label{sW-X}
 \eea
Here 
  $\bar{\D}$ denotes the chiral projection operator \cite{KT-M09,Muller}
\bea
\bar{\D}
&=&\frac{1}{96} \Big((\cDB^{ij}+16\bar{S}^{ij})\cDB_{ij}
-(\cDB^{\ad\bd}-16\bar{Y}^{\ad\bd})\cDB_{\ad\bd} \Big)
\non\\
&=&\frac{1}{96} \Big(\cDB_{ij}(\cDB^{ij}+16\bar{S}^{ij})
-\cDB_{\ad\bd}(\cDB^{\ad\bd}-16\bar{Y}^{\ad\bd}) \Big)~,
\label{chiral-pr}
\eea
with $\cDB^{\ad\bd}:=\cDB^{(\ad}_k\cDB^{\bd)k}$.
The  main properties of $\bar \D$ can be formulated using 
a super-Weyl inert scalar $U$ as follows:
\begin{subequations} 
\bea
{\bar \cD}^{\ad}_i \bar{\D} U &=&0~, \\
\d_\s U = 0 \quad \Longrightarrow \quad 
\d_\s \bar \D U &=& 2\s \bar \D U~,  \label{2.5b}\\
\int \rd^4 x \,{\rm d}^4\q\,{\rm d}^4{\bar \q}\,E\, U
&=& \int {\rm d}^4x \,{\rm d}^4 \q \, \cE \, \bar{\D} U ~.
\label{chiralproj1} 
\eea
\end{subequations}
Here $E$ denotes  the full superspace density. 

The super-Weyl invariance of \eqref{2.2} follows 
from the relations \eqref{sW-X} and  \eqref{chiralproj1} 
in conjunction with the identity 
\bea
{\bar \cD}^{\ad}_i  \s=0
\quad \Longrightarrow \quad 
\int \rd^4 x \,{\rm d}^4\q\,{\rm d}^4{\bar \q}\,E\, \s=0~,
\label{2.6}
\eea
for any covariantly chiral scalar $\s$. 

As shown in  \cite{BdeWKL},  the functional 
\bea
\int \rd^4x\, \rd^4\q\, \cE\, \Big\{ W^{\a\b}W_{\a\b}
-\X\Big\} 
\label{topological}
\eea
is a topological invariant being related to the difference 
of the Gauss-Bonnet and Pontryagin invariants. 

The specific feature of the invariants \eqref{2.1} and \eqref{2.2} is that 
they do not involve any conformal compensator.\footnote{Super-Weyl anomalies in 
$\cN=2$ superconformal theories coupled to supergravity \cite{Kuzenko:2013gva}
are given by linear combinations of the integrands 
in  \eqref{2.1} and \eqref{2.2}.}
However, such a compensator is required in order to construct 
a supersymmetric extension of the $R^2$ term,
and it should be the improved tensor multiplet 
\cite{deWPV}.

The tensor  (or {\it linear}) multiplet can be described in curved superspace by
its gauge invariant field strength $\cG^{ij}$  which is defined to be a  real ${\rm SU}(2)$ triplet 
subject to the covariant constraints  \cite{BS,SSW}
\bea
\cD^{(i}_\a \cG^{jk)} =  {\bar \cD}^{(i}_\ad \cG^{jk)} = 0~.
\label{1.2}
\eea
These constraints are solved in terms of a chiral
prepotential $\Psi$ \cite{HST,GS82,Siegel83,Muller86} via
\begin{align}
\label{eq_Gprepotential}
\cG^{ij} = \frac{1}{4}\Big( \cD^{ij} +4{S}^{ij}\Big) \Psi
+\frac{1}{4}\Big( \cDB^{ij} +4\bar{S}^{ij}\Big){\bar \Psi}~, \qquad
{\bar \cD}^i_\ad \J=0~,
\end{align}
which is invariant under shifts
$\Psi \rightarrow \Psi + \ri \Lambda$,
with $\Lambda$ a {\it reduced chiral} superfield,\footnote{The field strength 
$\cW$ of an Abelian vector multiplet is a reduced chiral superfield.}
\bea
\cDB^\ad_i \L= 0~, \qquad
\Big(\cD^{ij}+4S^{ij}\Big) \L &=&
\Big(\cDB^{ij}+4\bar{S}^{ij}\Big)\bar{\L} ~.
\eea
The super-Weyl transformation laws of $\J$ and $\cG^{ij}$ are 
\bea
\d_\s \J =   \s \J \quad \Longrightarrow \quad
\d_\s \cG^{ij} &=& ( \s +\bar \s)  \cG^{ij}~.
\eea
The improved tensor multiplet is characterised by the condition 
$\cG^2 := \hf \cG^{ij} \cG_{ij} \neq 0$.

Using the improved tensor multiplet one can construct the following 
reduced chiral superfield
$\mathbb W$:
\bea \mathbb W &=& - \frac{1}{24 \cG} (\bar \cD_{ij} + 12 \bar S_{ij}) \cG^{ij}
+ \frac{1}{36 \cG^3} \bar \cD_{\ad k} \cG^{ki} \bar \cD^\ad_l \cG^{lj} \cG_{ij} \non\\
&=& 
- \frac{\cG}{8} (\bar \cD_{ij} + 4 \bar S_{ij}) \frac{\cG^{ij}}{\cG^2} \ . \label{bbWSU(2)}
\eea
The regular procedure to derive $\mathbb W $ is described in \cite{BK11}.
This multiplet (up to normalisations) was discovered originally in \cite{deWPV}
using superconformal tensor calculus.
It was later reconstructed  in curved superspace
 \cite{Muller86} with the aid of the results in \cite{deWPV} and \cite{Siegel85}.

The supersymmetric completion of the $R^2$ term will prove to be described by the invariant
\begin{align}
I_{R^2} = \int \rd^4x\, \rd^4\q\, \cE\,  {\mathbb W}^2 + {\rm c.c.} \label{R2Action}
\end{align}
In the next section we will explicitly show that the above invariant does indeed contain a 
$R^2$ term at the component level.

The scalar curvature squared invariant \eqref{R2Action} is analogous to the one constructed in five dimensions in \cite{OP13}. There a supersymmetric completion of a $R^2$ term was obtained by considering the Chern-Simons coupling between a vector multiplet and two identical composite vector multiplets constructed out of the tensor multiplet. A similar procedure was performed in superspace in \cite{BKNT-M14}. In contrast to five dimensions the component action corresponding to \eqref{R2Action} contains a $R^2$ term without the need to impose any gauge condition.

%%%%%%%%%%%%%%%%%%%%%%%%%%%%%%%%%%%%%%%%%%%%%%%%%%%%%%

%%%%%%%%%%%%%%%%%%%%%%%%%%%%%%%%%%%%%%%%%%%%%%%%%%%%%%
%%%%%%%%%%%%%%%%%%%%%%%%%%%%%%%%%%%%%%%%%%%%%%%%%%%%%%

\section{Supersymmetric invariants in components} \label{section3}

The curvature squared invariants \eqref{2.1} and \eqref{2.2} are independent of 
any compensator and were reduced to components in \cite{BdeWKL}. In order to perform 
component reduction of the $R^2$ invariant \eqref{R2Action} 
it is advantageous to lift the superspace actions to conformal superspace.

Besides the $R^2$ invariant \eqref{R2Action} is it also worth elaborating on 
the component structure of the tensor multiplet action \cite{deWPV}, which in 
the formulation of \cite{KLRT-M1} is given by
\begin{align}
S_{\rm tensor} = -\int \rd^4x\, \rd^4\q\, \cE\, \Psi \mathbb W + {\rm c.c.} \ , \label{tensorAction}
\end{align}
where $\mathbb W$ is given in eq. \eqref{bbWSU(2)}. The above 
action can be shown to contain an Einstein-Hilbert term upon imposing a 
certain gauge.

In this section we first lift the descriptions of the tensor multiplet action and the 
$R^2$ invariant to conformal superspace and then reduce them to components. 
The salient details of conformal superspace are summarised in Appendix \ref{confSuperspace}.

%%%%%%%%%%%%%%%%%%%%%%%%%%%%%%%%%%%%%%%%%%%%%%%%%%%%%%

\subsection{Invariants in conformal superspace}

The tensor multiplet is described in conformal superspace by a real 
primary superfield $\cG^{ij} = \cG^{ji}$ of dimension $2$,
\be
\mathbb D \cG^{ij} = 2 \cG^{ij} \ , \quad K_A \cG^{ij} = 0
\ ,
\ee
satisfying the constraint
\be \nabla_\a^{(i} \cG^{jk)} = 0 \ .
\ee
The tensor multiplet can be described by a two-form gauge potential. 
Its superform formulation in conformal superspace 
can be found in \cite{BN12}. The 
tensor multiplet can be solved in terms of an unconstrained chiral prepotential $\Psi$,
\be \cG^{ij} = \frac{1}{4} \nabla^{ij} \Psi + \frac{1}{4} \bar \nabla^{ij} \bar\Psi \ ,
\ee
where we have defined $\nabla^{ij} := \nabla^{\a (i} \nabla_\a^{j)}$.

We can lift the superspace 
expressions for the tensor multiplet and the scalar curvature squared actions 
to conformal superspace. In conformal superspace, the tensor multiplet 
action is defined by \eqref{tensorAction}
but with the composite $\mathbb W$ now constructed with the 
covariant derivative of conformal superspace:
\bea \mathbb W &=& - \frac{1}{24 \cG} \bar \nabla_{ij} \cG^{ij}
+ \frac{1}{36 \cG^3} \bar \nabla_{\ad k} \cG^{ki} \bar \nabla^\ad_l \cG^{lj} \cG_{ij} = 
 - \frac{\cG}{8} \bar \nabla_{ij} \frac{\cG^{ij}}{\cG^2} \ . \label{bbWconfSuperspace}
\eea
One can check that $\mathbb W$ is indeed a vector multiplet since it is a  
primary superfield of dimension 1 satisfying the reduced chiral constraints
\be \bar \nabla_\ad^i \mathbb W = 0 \ , \quad \nabla^{ij} \mathbb W = \bar \nabla^{ij} \bar{\mathbb W} \ .
\ee
The above expression for $\mathbb W$ degauges to the one given by eq. \eqref{bbWSU(2)} upon using 
the degauging procedure given in \cite{Butter11}.

It is important to note that the action \eqref{tensorAction} only involves $\cG^{ij}$ without 
a compensating vector multiplet. This is in contrast to, for example, 
the $\cN=2$ supersymmetric $BF$ action\footnote{In general, a $BF$ theory on a 
$d$-dimensional orientable manifold
is a Schwarz-type topological gauge theory with action 
$S_{(d,n)} =\int  B_n \wedge \rd A_{d-n-1}= \int  B_n \wedge F_{d-n}$, 
where $B_n$ and $ A_{d-n-1} $ are differential forms and $F_{d-n}$ is the gauge invariant
field strength associated with $A_{d-n-1}$. For a review  of $BF$ theories, see \cite{BBRT}.
The action \eqref{PsiW} is a supersymmetric generalisation of $S_{(4,2)}$.} 
which is described by
\begin{align}\label{PsiW}
S_{BF} = \int \rd^4x\, \rd^4\q\, \cE\, \Psi \cW + {\rm c.c.} \ ,
\end{align}
where $\cW$ is the chiral field strength of a  vector multiplet.

In conformal superspace, the scalar curvature squared action is described by eq. \eqref{R2Action} 
but with the composite $\mathbb W$ replaced by the one given in 
eq. \eqref{bbWconfSuperspace}.
It is important to emphasise that the invariant \eqref{R2Action} 
has the same type as 
the vector multiplet action,
\be S_{\rm vector} = \int \rd^4x\, \rd^4\q\, \cE\, \cW^2 + {\rm c.c.} \label{VMaction}
\ee
with the vector superfield strength replaced by the composite $\mathbb W$. 
The 
component vector multiplet and tensor multiplet actions
were given in \cite{BN12} in our notation and conventions.\footnote{However, here we denote the 
Lorentz curvature constructed from the spin connection by $R_{ab}{}^{cd}$ 
instead of $\hat \cR_{ab}{}^{cd}$.} In the next subsection we apply the 
results of \cite{BN12} to elaborate on the component structure of the invariants $S_{\rm tensor}$ and 
$I_{R^2}$. 

%%%%%%%%%%%%%%%%%%%%%%%%%%%%%%%%%%%%%%%%%%%%%%%%%%%%%%

\subsection{The tensor multiplet action and scalar curvature squared invariant in components}

Here we identify the component fields of 
the Weyl multiplet of conformal supergravity
in accordance with 
\cite{Butter11}. 
The vierbein $e_m{}^a$, 
the gravitino $\psi_m{}^\a_i$, the $\rm{ U}(1)$ and $\rm{ SU}(2)$
gauge fields $A_m$ and $\phi_m{}^{ij}$, and the dilatation gauge field $b_m$
are 
defined as follows:
\begin{align}
e_m{}^a &:= E_m{}^a| \ , \quad \psi_m{}^\a_i := 2 E_m{}^\a_i| \ , \quad {\bar \psi}_m{}^i_\ad := 2 E_m{}^i_\ad| \ , \non\\
A_m &:= \Phi_m| \ , \quad \phi_m{}^{i j} := \Phi_m{}^{ij}| \ ,\quad b_m := B_m| ~.
\end{align}
The component (or bar) projection of a superfield $V(z)$ is defined in the usual way 
$V| := V(z)|_{\theta = \bar{\theta} = 0}$. There are several 
\emph{composite} gauge connections.  
The spin connection  $\omega_m{}^{ab}$, the special
conformal $\frak{f}_m{}^a$
and $S$-supersymmetry connections $\phi_m{}_\a^i$,
\begin{align}
\omega_m{}^{ab} := \Omega_m{}^{ab}| \ , \quad  
\frak{f}_m{}^a := \frak{F}_m{}^a| \ , \quad
\phi_m{}^i_\a := 2 \frak F_m{}^i_\a| \ ,
\end{align}
are all composed of the previously defined component fields. Their expressions can be found 
in \cite{Butter11, BN12}.

The Weyl multiplet also contains some  non-gauge fields. 
These are encoded in the components of $W_{\a\b}$ 
as follows:
\begin{subequations}
\begin{gather}
W_{ab} = W_{ab}^{+} + W_{ab}^{-} \ , \quad W_{ab}^{+} := (\s_{ab})^{\a\b} W_{\a\b}| \ , \quad W_{ab}^{-} := - (\tilde{\s}_{ab})_{\ad\bd} \bar{W}^{\ad\bd}| \ ,\\
\S^{\a i} := \frac{1}{3} \nabla_\b^i W^{\a\b}| \ , \quad
D := \frac{1}{12} \nabla^{\a\b} W_{\a\b}| = \frac{1}{12} \bar{\nabla}_{\ad\bd} \bar{W}^{\ad\bd}| \ .
\end{gather}
\end{subequations}
The component field $W_{ab}^{\pm}$ satisfies the self-duality relation
$\frac{\ri}{2} \eps_{ab}{}^{cd} W_{cd}^{\pm} = \pm W_{ab}^{\pm} $. As in \cite{BN12}, to avoid 
cluttered notation, we will often use $W_{\a\b}$ also for the corresponding component field. It 
should be clear from context to which we are referring. In what follows, we will also make use of the 
following bosonic 
covariant derivative:
\be \nabla'_a = e_a{}^m \Big(\partial_m
	+ \hf \omega_m{}^{bc} M_{bc}+ \phi_m{}^{ij} J_{ij} + \ri A_m Y + b_m \mathbb{D}  \Big)~.
\ee

The matter components of the tensor multiplet are defined as follows:
\be
G^{ij} := \cG^{ij}|~, \quad 
\c_{\a i} := \frac{1}{3} \nabla_\a^j \cG_{i j}| ~, 
\quad 
F :=  \frac{1}{12} \nabla^{ij} \cG_{ij}|~.
\ee
There is also an additional component field, the two-form $b_{mn}$. Its supercovariant field 
strength is given by
\bea
\tilde{h}^a &=& \frac{\ri}{24} (\s^d)^\a{}_\bd [\nabla_\a^i , \bar \nabla^\bd_j] \cG^j{}_i| \non\\
	&=& \hf \ve^{abcd} (\frac{1}{3} h_{bcd}
	- \ri (\s_{cd})_\a{}^\b \psi_b{}^\a_k \chi_\b^k
	- \ri (\tilde{\s}_{cd})^\ad{}_\bd {\bar \psi}_b{}^k_\ad \bar{\chi}^\bd_k
	+  (\s_b)_\a{}^\bd \psi_c{}^\a_k {\bar \psi}_d{}^l_\bd G^k{}_l ) \ ,
\eea
where
\be h_{abc} = 3 e_a{}^m e_b{}^n e_c{}^p \partial_{[m} b_{np]} \ . 
\ee

The tensor multiplet action 
was reduced from conformal superspace to components  
 in \cite{BN12}. The action up to fermion contributions is
\bea \label{compTensor}
S_{\rm tensor} &=&
     \int \rd^4x \, e \,\Big(  
      \frac{1}{2G} |F|^2 
      - G \Big(\frac{1}{3} R + D \Big)
           -\frac{1}{2G} \Big( 
           \tilde{h}^a \tilde{h}_a
      - G_{ij} \nabla'_a \nabla'^a G^{ij}
      \Big) \non\\
      &&- \frac{1}{4 G^3} G_{ij} \nabla'{}^a G^{ik} \nabla'_a G^{jl} G_{kl}
	+\frac{1}{2} \eps^{mnpq} b_{mn} f_{pq} 
	\Big) + {\rm fermion \ terms} \ ,
\eea
where
\be f_{mn} = 2 \partial_{[m} \G_{n]} + \frac{1}{4 G^3} \partial_m G^{ik} \partial_n G_k{}^j G_{ij} \ , \quad 
G = \cG|
\ee
and
\be
\Gamma_m := 
	\frac{1}{2G} \phi_{m}{}^{ij} G_{ij} + \frac{1}{2G} e_{m}{}^a \tilde h_a + {\rm fermion \ terms} \ .
\ee
We see that in the gauge where $G = 1$, $S_{\rm tensor}$ 
contains the Einstein-Hilbert term with a wrong sign.

The component form for the invariant $I_{R^2}$
may be obtained 
by replacing the component fields of the vector multiplet in the vector multiplet action 
in \cite{BN12} with the component fields of the composite $\mathbb W$. 
The invariant  up to 
fermionic contributions is
\bea \label{compIR2}
I_{R^2} &=&
     \int \rd^4x \, e \,\Big( - \nabla'^a (G^{-1} F) \nabla_a' (G^{-1} \bar F)
     + \frac{1}{G^2} |F|^2 D
	- \frac{1}{6 G^2} R |F|^2 \non\\
     &&+ \frac{1}{8} X^{ij} X_{ij}
     - 2 f^{ab} f_{ab}
     - \frac{2}{G}(\s^{ab})_{\ad\bd} \bar{F} \bar W^{\ad\bd} f_{ab}
     + \frac{2}{G} (\s^{ab})_{\a\b} F W^{\a\b} f_{ab} \non\\
     &&- \frac{1}{2 G^2} \bar W_{\dalpha \dbeta} \bar W^{\dalpha \dbeta} \bar F^2
     - \frac{1}{2 G^2} \, W^{\alpha \beta} W_{\alpha \beta} F^2 \Big) + {\rm fermion \ terms} \ ,
\eea
where
\bea
X^{ij} &=& \frac{1}{G} \Big(- 2 \nabla'_a \nabla'^a G^{ij} + 2 D G^{ij} + \frac{2}{3} R G^{ij}  \Big) 
+ \frac{1}{G^3} \Big( \nabla'^a G^{ik} \nabla'_a G^{jl} G_{kl} \non\\
&&+ \tilde{h}^a \tilde{h}_a G^{ij} + G^{ij} F \bar{F} 
- 2 \tilde{h}^a \nabla'_a G^{k(i} G^{j)}{}_k \Big) + {\rm fermion \ terms}
\eea
and
\be f_{ab} = e_a{}^m e_b{}^n f_{mn} \ .
\ee
One can see that a $R^2$ term arises in the $X^{ij} X_{ij}$ contribution to the 
invariant \eqref{compIR2}. It should be emphasised that the invariant \eqref{compIR2} is
 independent of any gauge choice.

%%%%%%%%%%%%%%%%%%%%%%%%%%%%%%%%%%%%%%%%%%%%%%%%%%%%%%
%%%%%%%%%%%%%%%%%%%%%%%%%%%%%%%%%%%%%%%%%%%%%%%%%%%%%%

\section{Discussion} \label{section4}

In this paper we have completed the description of all off-shell curvature squared 
invariants in $\cN = 2$ supergravity. 
Such
invariants are described by the linear combination
\be 
I = A I_{C^2_{abcd}} +B I_{R^2_{ab} -\frac{1}{3} R^2} + C I_{R^2}\ ,
\ee
where $A$, $B$ and $C$ are real parameters, 
and the invariants $I_{C^2_{abcd}}$, $I_{R^2_{ab} -\frac{1}{3} R^2} $ and $ I_{R^2}$
are given by the equations \eqref{2.1}, \eqref{2.2} and \eqref{R2Action}, 
respectively. 
The bosonic sector of $I_{R^2} = \int \rd^4 x \, e \, L_{R^2}$ requires some discussion.

First of all, let us consider the part of $ L_{R^2} $ containing the auxiliary scalar $D$:
\bea
L_{R^2} &=&
     - \frac{D}{2 G^2} \Big( - 4 \nabla'^a G \nabla'_a G + \nabla'{}^{a} G^{ij} \nabla'_a G_{ij}
     - 2 \tilde{h}^a \tilde{h}_a \non\\
     &&+ 2 \nabla'^a \nabla'_a G^{ij} G_{ij} 
      - \frac{4}{3} R G^2 - 4 |F|^2
     \Big) + D^2 + \cdots 
     \ ,
     \label{4.2}
\eea
where the ellipsis represents terms not directly involving $D$.
We see that the equation of motion for $D$ is consistent and
allows one to integrate $D$ out.  To understand the importance of this result, 
it is worth recalling why two compensators are required in 
ordinary $\cN=2$ supergravity \cite{deWPV}. 
The $D$-terms in the vector and tensor multiplet
Lagrangians are
\bea
L_{\rm vector} &=&
     4 | \phi |^2 D+ \cdots 
     \ ,\qquad L_{\rm tensor} =
     - D G + \cdots ~,
\eea
where $\phi = \cW|$.
It is seen that the equation of motion for $D$ is contradictory if one chooses 
$L_{\rm SG} = - L_{\rm vector} $ or $L_{\rm SG} = - L_{\rm tensor} $.
However, the supergravity Lagrangian $L_{\rm SG} = - L_{\rm vector} - L_{\rm tensor} $
leads to a sensible equation of motion for $D$, which is $G= 4 |\f |^2$. 
Now, looking at the Lagrangian \eqref{4.2}, we see that 
we can circumvent the need of having two compensators. We can 
have the invariant $I_{R^2}$ on its own or we can 
have a linear combination of $I_{R^2}$ and 
$S_{\rm tensor}$, and still have consistent 
dynamics.

The second important feature of $I_{R^2}$ concerns the component field 
$W^{\a\b}$ and its conjugate, which are auxiliary in ordinary $\cN=2$ supergravity.
The equation of motion for $W^{\a\b}$, which is derived from $I_{R^2}$, allows one 
to integrate $W_{\a\b} $ out (if we assume $F \neq 0$) giving
\bea
I_{R^2} &=&  
     \int \rd^4x \, e \,\Big( - \nabla'^a (G^{-1} F) \nabla_a' (G^{-1} \bar F)
     + \frac{1}{G^2} |F|^2 D\non\\
	&&- \frac{1}{6 G^2} R |F|^2 
     + \frac{1}{8} X^{ij} X_{ij}
     + 2 f^{ab} f_{ab}
     + {\rm fermion \ terms} \Big) \ .
\eea
Comparing with eq. \eqref{compIR2}, one notices that there is a 
sign flip in the $f^{ab} f_{ab}$ term. In the above action the 
auxiliary field has not yet been eliminated. In ordinary supergravity $F$ was 
auxiliary, however, now it becomes dynamical. It should also be noted 
that upon eliminating the auxiliary field $D$ in the combined invariant of the form
$S_{\rm tensor} + I$,
we generate not only an $R^2$ term but also an $R$ and a potential $|{F/G}|^4$ term. 
This will be discussed below.

Let us further elaborate on the elimination of the auxiliary field $D$. To begin with 
we consider the invariant
\be \a S_{\rm vector} + \b S_{\rm tensor} + \g I_{R^2} \ , \label{invariantR^2+tensor}
\ee
where $\a$, $\b$ and $\g$ are arbitrary constants. The bosonic component action can be 
constructed by using \eqref{compTensor} and \eqref{compIR2}, and the vector multiplet action
in \cite{BN12}. 
One can check that upon eliminating the auxiliary field $D$, the $R^2$ term 
cancels.\footnote{We 
are grateful to Sergey Ketov for pointing out this observation. \label{KetovAck}} 
It is worth mentioning that in five dimensions there also exists a $R^2$ invariant in the standard Weyl 
multiplet \cite{OP13, BKNT-M14}, which involves an auxiliary 
scalar field.\footnote{A $\r R^2$ invariant in the standard Weyl multiplet 
was given in \cite{OP13} since a different gauge condition was used.} 
Upon eliminating the auxiliary field, the $R^2$ term can be similarly shown to vanish.

The component action of \eqref{invariantR^2+tensor} will  
contain the potential contribution
\be
- \frac{1}{\g}\Big(\frac{\a}{2} G - 2 \b |\phi|^2\Big)^2 \ ,
\ee
as well as the term
\be 
- \frac{4}{3} \b R |\phi|^2 \ .
\ee
Upon imposing an appropriate gauge, the first term contains a cosmological term, while 
the second gives rise to an Einstein-Hilbert term in the gauge $\phi = \rm const$.

Although the invariant $I_{R^2}$ does not give rise to a pure $R^2$ term alone upon integrating out the 
auxiliary field $D$, it can still lead to a non-trivial $R^2$ contribution if one adds to it another invariant. 
For instance, one can make use of the invariant \eqref{2.2} and consider the linear combination
\be 
 \cA S_{\rm tensor} + \cB I_{R^2} + \cC I_{R^2_{ab} -\frac{1}{3} R^2} \ .
\ee
The invariant $I_{R^2_{ab} -\frac{1}{3} R^2}$ contains a $D^2$ term 
and no $R D$ term at the component level, 
see \cite{BdeWKL}. This allows one to keep a $R^2$ contribution from $I_{R^2}$ upon eliminating the auxiliary field $D$. The 
invariant will also obtain an Einstein-Hilbert term and a cosmological constant 
in a gauge where $G = 1$.

It should be mentioned that one can fix the special conformal transformations, 
dilatations and break the $\rm SU(2)$ R-symmetry down 
to $\rm U(1)$ by imposing the following gauge conditions on the improved tensor multiplet
\be B_A = 0 \ , \quad \cG^{ij} = \d^{ij} \cG \ , \quad \cG = 1 \ . \label{gaugeCond}
\ee
These conditions correspond to the following choice for the component fields:
\be b_m = 0 \ , \quad G = 1 \ , \quad \c_\a^i = \bar \c_\ad^i = 0 \ , \quad G_{ij} = \d_{ij} G \ .
\ee
The first gauge choice fixes the special conformal summetry, the second
fixes dilatations, the third fixes the $S$-supersymmetry 
transformations and the last breaks the $\rm SU(2)$ R-symmetry to $\rm U(1)$. Upon imposing 
the above gauge choice, the $R^2$ invariant \eqref{R2Action} coincides with the $R^2$ invariant 
in \cite{Ketov}, which was only specified in the above gauge in superspace. Our invariant, however, is described in 
both conformal superspace and the conventional superspace formulation of \cite{KLRT-M1} without 
specifying any gauge condition on the compensator. As demonstrated, it is also readily reduced to components using the 
results of \cite{BN12}.

It may be shown that no $R^2$ invariant can be constructed with a 
compensating vector multiplet 
only. However, 
one can generalise the invariant $I_{R^2}$ by coupling it to some vector multiplets. 
One can consider the following simple generalisation 
of $I_{R^2}$:
\be 
\int \rd^4x\, \rd^4\q\, \cE\,  \cF + {\rm c.c.} \ ,
\ee
where $\cF$ is a homogeneous function of degree two in $\cW^{\hat I} = (\mathbb W, \cW^I)$ with 
$\cW^I$ denoting a number of vector multiplets. Such invariants were considered only in the 
gauge \eqref{gaugeCond} in \cite{Ketov}. Using the results of \cite{BN12} one can reduce the 
action to components.

So far we have considered constructing a $R^2$ invariant using a single compensator. Is it 
possible to use both a compensating tensor and vector multiplet together to generate a $R^2$ term? 
In principle, 
other invariants may be constructed with the use of the 
results in \cite{BK11}.  For instance, one can consider a projective-superspace
Lagrangian $\cL^{++} (v) $ of the form 
\bea 
\cL^{++}  =  \frac{(H^{++})^2}{G^{++}} \ , \qquad H^{++} (v) = H^{ij} v_i v_j
\ , \qquad G^{++} (v) = G^{ij} v_i v_j~,
\eea
where $H^{ij} = \nabla^{ij} \cW$ and  $v^i \in {\mathbb C}^2 \setminus \{0\}$ denotes
homogeneous coordinates for ${\mathbb C}P^1$. Using the results of \cite{BK11}, the corresponding invariant may be 
cast in the form of a $BF$ action, 
$\int \rd^4x\, \rd^4\q\, \cE\, \Psi \mathbb W_2 + {\rm c.c.}$, where
\be 
\mathbb W_2 
= - \frac{\cG}{16} \bar \nabla_{ij} \Big\{ \frac{1}{\cG^4} 
\Big( \d^i_{(k} \d^j_{l)}  -\frac{1}{2\cG^2} \cG^{ij} \cG_{kl}\Big)
H^{(kl} H^{mn)} \cG_{mn} \Big\} \ .
\ee
It can be shown that the invariant contains a $\frac{|\phi|^2}{G} R^2$ term at the component level. 
In pure supergravity, ${|\phi|^2}={G} $ on the mass shell, and then the above invariant 
gives a $R^2$ term. However, this condition does not hold in general. 
We conclude that our $R^2$ invariant   \eqref{R2Action} has no obvious alternative.

%%%%%%%%%%%%%%%%%%%%%%%%%%%%%%%%%%%%%%%%%%%%%%%%%%%%%%
%%%%%%%%%%%%%%%%%%%%%%%%%%%%%%%%%%%%%%%%%%%%%%%%%%%%%%
%%%%%%%%%%%%%%%%%%%%%%%%%%%%%%%%%%%%%%%%%%%%%%%%%%%%%%
$~$\\
\noindent
{\bf Acknowledgements:}\\
We are grateful to Stefan Theisen for asking a question that lead to the 
construction described in this paper. 
We also acknowledge Sergey Ketov for useful discussions and for ponting out an error 
in the first version of this paper (see footnote \ref{KetovAck}).
This work was supported by the ARC projects DP140103925 and DE120101498.

%%%%%%%%%%%%%%%%%%%%%%%%%%%%%%%%%%%%%%%%%%%%%%%%%%%%%%
%%%%%%%%%%%%%%%%%%%%%%%%%%%%%%%%%%%%%%%%%%%%%%%%%%%%%%

\appendix

%%%%%%%%%%%%%%%%%%%%%%%%%%%%%%%%%%%%%%%%%%%%%%%%%%%%%%
%%%%%%%%%%%%%%%%%%%%%%%%%%%%%%%%%%%%%%%%%%%%%%%%%%%%%%

\section{$\rm SU(2)$ superspace} \label{SU(2)geometry}
\label{AppendA}

This appendix contains a brief summary of the formulation for $\cN = 2$ conformal 
supergravity \cite{KLRT-M1} in SU(2) superspace \cite{Grimm}. Our notation and conventions 
follow those of \cite{Ideas}.

To describe SU(2) superspace one begins with a curved $\cN = 2$ superspace $\cM^{4|8}$ 
parametrised by
local 
coordinates 
$z^M = (x^m, \ \q^\mu_\imath, \ \bar{\theta}_{\dot{\mu}}^\imath 
= (\theta_{\mu \imath})^*)$, where
$m = 0, 1, \cdots, 3,$ $\mu = 1, 2$, $\dot{\mu} = 1, 2$ and $\imath = \1, \2$. 
The structure group is chosen to be $\rm SL(2, \dsC) \times SU(2)$, 
and the covariant derivatives $\cD_A = (\cD_a, \cD_\a^i, \bar \cD^\ad_i)$ 
have the form
\bea
\cD_A &=& E_A + \Phi_A{}^{kl} J_{kl}+ \hf \Omega_A{}^{bc} M_{bc} \non \\
		  &=& E_A + \Phi_A{}^{kl} J_{kl}+ \Omega_A{}^{\b\g} M_{\b\g} 
		  + \bar{\Omega}_A{}^{ \dot{\b} \dot{\g} } \bar M_{\dot{\b}\dot{\g}}~.
\eea
Here $E_A = E_A{}^M(z) \partial_M$ is the supervielbein, with $\partial_M = \partial/\partial z^M$,
$J_{kl} = J_{lk}$ are generators of the group $\rm SU(2)$ and
$M_{ab}$ are the Lorentz generators. The one-forms $\Omega_A{}^{bc}$ and $\Phi_A{}^{kl}$ 
are the Lorentz and SU(2) connections.

The generators act on the covariant derivatives as follows:
\be
[M_{\a\b},  \cD_\g^i] = \ve_{\g (\a } \cD_{\b)}^i \ , \qquad 
\big[ J_{kl}, \cD_\a^i \big] = - \d^i_{(k} \cD_{\a l)} \ .
\ee

The algebra of covariant derivatives is \cite{KLRT-M1}
\begin{subequations} \label{A.3}
\bea
\{\cD_\a^i,\cD_\b^j\}&=&\phantom{+}
4S^{ij}M_{\a\b}
+2\ve^{ij}\ve_{\a\b}Y^{\g\d}M_{\g\d}
+2\ve^{ij}\ve_{\a\b}\bar{W}^{\gd\dd}\bar{M}_{\gd\dd}
\non\\
&&
+2 \ve_{\a\b}\ve^{ij}S^{kl}J_{kl}
+4 Y_{\a\b}J^{ij}~,
\label{acr1} \\
\{\cD_\a^i,\cDB^\bd_j\}&=&
-2\ri\d^i_j(\s^c)_\a{}^\bd\cD_c
+4\d^{i}_{j}G^{\d\bd}M_{\a\d}
+4\d^{i}_{j}G_{\a\gd}\bar{M}^{\gd\bd}
+8 G_\a{}^\bd J^{i}{}_{j}~.
\label{acr3} 
\eea
\end{subequations}
The explicit expressions for the commutator
${[}\cD_a,\cD_\b^j{]}$ can be found in \cite{KLRT-M1}. 
Here the real four-vector $G_{\a \ad} $,
the complex symmetric  tensors $S^{ij}=S^{ji}$, $W_{\a\b}=W_{\b\a}$, 
$Y_{\a\b}=Y_{\b\a}$ and their complex conjugates 
$\bar{S}_{ij}:=\overline{S^{ij}}$, $\bar{W}_{\ad\bd}:=\overline{W_{\a\b}}$,
$\bar{Y}_{\ad\bd}:=\overline{Y_{\a\b}}$ 
are constrained by 
certain Bianchi identities \cite{Grimm,KLRT-M1}.
 The latter  comprise 
 the  dimension-3/2 identities 
\begin{subequations}\label{A.5}
\bea
\cD_{\a}^{(i}S^{jk)}= {\bar \cD}_{\ad}^{(i}S^{jk)}&=&0~,
\label{S-analit}\\
\cDB^\ad_iW_{\b\g}&=&0~,\qquad\\
\cD_{(\a}^{i}Y_{\b\g)}&=&0~,\\
\cD_{\a}^{i}S_{ij}+\cD^{\b}_{j}Y_{\b\a}&=&0~, \\
\cD_\a^iG_{\b\bd}&=&
-{\frac14}\cDB_{\bd}^iY_{\a\b}
+\frac{1}{12}\ve_{\a\b}\cDB_{\bd j}S^{ij}
-{\frac14}\ve_{\a\b}\cDB^{\gd i}\bar{W}_{\bd\gd}~,
\eea
\end{subequations}
as well as the dimension-2 relation
\bea
\big( \cD^i_{(\a} \cD_{\b) i}
-4Y_{\a\b} \big) W^{\a\b}
&=& \big( \cDB_i^{( \ad }\cDB^{\bd ) i}
-4\bar{Y}^{\ad\bd} \big) \bar{W}_{\ad\bd}
~.
\label{dim-2-constr}
\eea

The algebra of covariant derivatives  \eqref{A.3} is invariant 
under the super-Weyl transformations  \cite{KLRT-M1}
\begin{subequations}
\bea
\d_{\s} \cD_\a^i&=&\hf\sba\cD_\a^i+(\cD^{\g i}\s)M_{\g\a}-(\cD_{\a k}\s)J^{ki}~, \\
\d_{\s} \cD_a&=&
\hf(\s+\sba)\cD_a
+{\frac{\ri}4}(\s_a)^\a{}_{\bd}(\cD_{\a}^{ k}\s)\cDB^{\bd}_{ k}
+{\frac{\ri}4}(\s_a)^{\a}{}_\bd(\cDB^{\bd}_{ k}\sba)\cD_{\a}^{ k} \non \\
&&\quad -{\frac12}\big(\cD^b(\s+\sba)\big)M_{ab}
~, 
\eea
\end{subequations}
with  the parameter $\s$ being an arbitrary covariantly chiral superfield, 
\bea
{\bar \cD}_{\ad i} \s=0~,
\eea
provided 
the  dimension 1 components of the torsion transform 
 as follows:
\begin{subequations}
\bea
\d_{\s} S^{ij}&=&\sba S^{ij}-{\frac14}\cD^{\g(i}\cD^{j)}_\g \s~, 
\label{super-Weyl-S} \\
\d_{\s} Y_{\a\b}&=&\sba Y_{\a\b}-{\frac14}\cD^{k}_{(\a}\cD_{\b)k}\s~,
\label{super-Weyl-Y} \\
\d_{\s} {W}_{\a \b}&=&\s {W}_{\a \b }~, \label{A.9c}\\
\d_{\s} G_{\a\bd} &=&
\hf(\s+\sba)G_{\a\bd} -{\frac{\ri}4}
\cD_{\a \bd} (\s-\sba)~.
 \label{super-Weyl-G} 
\eea
\end{subequations}
As is seen from \eqref{A.9c}, 
the super-Weyl tensor $W_{\a\b}$ transforms homogeneously.

%%%%%%%%%%%%%%%%%%%%%%%%%%%%%%%%%%%%%%%%%%%%%%%%%%%%%%
%%%%%%%%%%%%%%%%%%%%%%%%%%%%%%%%%%%%%%%%%%%%%%%%%%%%%%

\section{Conformal superspace} \label{confSuperspace}

In this appendix we present the salient details of the superspace formulation of 
$\cN = 2$ conformal supergravity \cite{Butter11}, known as conformal superspace. 
The SU(2) superspace of the previous section may be viewed as a gauged fixed version 
of conformal superspace \cite{Butter11}, which gauges the entire superconformal group 
$\rm SU(2, 2|2)$. Our conventions and presentation follows that of \cite{BN12}.

The covariant derivatives of conformal superspace
$\nabla_A = (\nabla_a, \nabla_\a^i , \bar{\nabla}^\ad_i)$ have the form
\be \nabla_A = E_A + \hf \Omega_A{}^{ab} M_{ab} + \Phi_A{}^{ij} J_{ij} + \ri \Phi_A Y 
+ B_A \mathbb{D} + \frak{F}_{A}{}^B K_B \ .
\ee
Here $Y$ is the generator of 
the U(1) subgroup of the $\cN=2$ R-symmetry group $\rm SU(2) \times U(1)$,
%the chiral rotation group $\rm U(1)_R$, 
and $K^A = (K^a, S^\a_i, \bar{S}_\ad^i)$ are the special
superconformal generators, while
the one-forms $\Phi_A$, $B_A$ and
$\frak{F}_A{}^B$ are the corresponding connections.

The Lorentz, $\rm SU(2)$, $\rm U(1)$ and dilatation generators 
act on the spinor covariant derivatives as
\bsubeq
\bea
[M_{ab}, \nabla_\a^i] &=& (\s_{ab})_\a{}^\b \nabla_\b^i ~,\quad
[J_{ij}, \nabla_\a^k] = - \d^k_{(i} \nabla_{\a j)} ~, \\
\left[ Y, \nabla_\a^i \right] &=& \nabla_\a^i ~,\quad
[\mathbb{D}, \nabla_\a^i] = \hf \nabla_\a^i ~.
\eea
\esubeq
The $S$-supersymmetry generators $S^\a_i$ transform under Lorentz, $\rm SU(2)$, 
$\rm U(1)_R$ and dilatations as
\bsubeq
\bea
[M_{ab} , S^\g_i] &=& - (\s_{ab})_\b{}^\g S^\b_i ~, \quad
[J_{ij}, S^\g_k] = - \ve_{k (i} S^\g_{j)} ~, \quad \\
\left[ Y, S^\a_i \right] &=& - S^\a_i ~, \quad
[\mathbb{D}, S^\a_i] = - \hf S^\a_i ~.
\eea
\esubeq
Among themselves, the generators $K^A$ obey the algebra
\begin{align}
\{ S^\a_i , \bar{S}^j_\ad \} &= 2 \ri \d^j_i (\s^a)^\a{}_\ad K_a~,
\end{align}
while the special conformal generators $K^A$ act on $\nabla_B$ as
\bsubeq
\bea
\{ S^\a_i , \nabla_\b^j \} &=& 2 \d^j_i \d^\a_\b \mathbb{D} - 4 \d^j_i M^\a{}_\b 
- \d^j_i \d^\a_\b Y + 4 \d^\a_\b J_i{}^j \ , \quad
[S^\a_i , \nabla_b] = \ri (\s_b)^\a{}_\bd \bar{\nabla}^\bd_i \ , ~~~~~\\
\left[ K^a, \nabla_\b^j \right] &=& -\ri (\s^a)_\b{}^\bd \bar{S}_\bd^j \ , \quad
\left[ K^a, \nabla_b \right] = 2 \delta^a_b \mathbb{D} + 2 M^{a}{}_b ~.
\eea
\esubeq

The algebra of covariant derivatives is
\begin{subequations}\label{CSGAlgebra}
\begin{align}
\{ \nabla_\a^i , \nabla_\b^j \} &= 2 \ve^{ij} \ve_{\a\b} \bar{W}_{\gd\dd} \bar{M}^{\gd\dd} + \hf \ve^{ij} \ve_{\a\b} \bar{\nabla}_{\gd k} \bar{W}^{\gd\dd} \bar{S}^k_\dd - \hf \ve^{ij} \ve_{\a\b} \nabla_{\g\dd} \bar{W}^\dd{}_\gd K^{\g \gd}~, \\
\{ \nabla_\a^i , \bar{\nabla}^\bd_j \} &= - 2 \ri \d_j^i \nabla_\a{}^\bd~, \\
[\nabla_{\a\ad} , \nabla_\b^i ] &= - \ri \ve_{\a\b} \bar{W}_{\ad\bd} \bar{\nabla}^{\bd i} - \frac{\ri}{2} \ve_{\a\b} \bar{\nabla}^{\bd i} \bar{W}_{\ad\bd} \mathbb{D} - \frac{\ri}{4} \ve_{\a\b} \bar{\nabla}^{\bd i} \bar{W}_{\ad\bd} Y + \ri \ve_{\a\b} \bar{\nabla}^\bd_j \bar{W}_{\ad\bd} J^{ij}
	\eol & \quad
	- \ri \ve_{\a\b} \bar{\nabla}_\bd^i \bar{W}_{\gd\ad} \bar{M}^{\bd \gd} - \frac{\ri}{4} \ve_{\a\b} \bar{\nabla}_\ad^i \bar{\nabla}^\bd_k \bar{W}_{\bd\gd} \bar{S}^{\gd k} + \frac{1}{2} \ve_{\a\b} \nabla^{\g \bd} \bar{W}_{\ad\bd} S^i_\g
	\eol & \quad
	+ \frac{\ri}{4} \ve_{\a\b} \bar{\nabla}_\ad^i \nabla^\g{}_\gd \bar{W}^{\gd \bd} K_{\g \bd}~.
\end{align}
\end{subequations}
The super-Weyl tensor $W_{\a\b} = W_{\b\a}$ and its complex conjugate
${\bar{W}}_{\ad \bd} := \overline{W_{\a\b}}$ are superconformally primary,
$K_A W_{\a\b} = 0$, and obey the additional constraints
\begin{align}
\bar{\nabla}^\ad_i W_{\b\g} = 0~,\qquad
\nabla_{\a}^k \nabla_{\b k} W^{\a\b} &= \bar\nabla^{\ad}_k \bar\nabla^{\bd k} \bar{W}_{\ad\bd} ~.
\end{align}

In contrast to SU(2) superspace the entire algebra of covariant derivatives is 
constructed in terms of the super-Weyl tensor $W_{\a\b}$.

%%%%%%%%%%%%%%%%%%%%%%%%%%%%%%%%%%%%%%%%%%%%%%%%%%%%%%
%%%%%%%%%%%%%%%%%%%%%%%%%%%%%%%%%%%%%%%%%%%%%%%%%%%%%%

\begin{footnotesize}

\end{footnotesize}

\end{document}